\documentclass[]{article}
\usepackage[english]{babel}
\usepackage{graphicx,enumerate}
\usepackage[tbtags]{amsmath}

\begin{document}
\title{Simple model of first-order phase transition}
\author{Marcin Ostrowski\footnote{e-mail: mostrowski@ahe.lodz.pl}}
\maketitle

\begin{abstract}
One-dimensional model of a system where first-order phase transition occurs is examined in the present paper. It is shown that basic properties
of the phenomenon, such as a well defined temperature of transition, are caused both by existence of a border between the phases and the fact
that only in the vicinity of that border it is possible for molecules to change their phase. Not only the model is introduced and theoretical
analysis of its properties is made but also the  results of Monte Carlo simulations are presented together with the results of numerical calculation
of the distribution of energy levels of the system.
\end{abstract}

\section{Introduction}
The prime purpose of the paper is to introduce a simple model of the system where first-order phase transition occurs. One of possible scenarios is
presented here, which may lead to such phenomena as melting-freezing or evaporation-condensation. The task is mainly of didactic character. It does
not discover any new types of systems in question but, basing on a very simple model, shows what are essential properties of the first-order
transition. The model is meant to provide an introductory exercise for those students who plan to get involved in research of such phenomena, that is
why numerous simplifications are applied in the model.\\
Due to simplicity, our model allows for formal (mathematical) calculation of some thermodynamic quantities (e.g. temperature of the transition, latent
heat per molecule) as well as for the use of numerical simulation methods. Both approaches were taken in the present work.

It is well known that one of the simplest attempts to classify phase transitions takes into account existence of  latent heat. Hence, first-order phase
transitions are those that involve latent heat (e.g. melting, evaporation) while second-order phase transitions do not (e.g. ferromagnetic to
paramagnetic transition).

There are a lot of models which describe phase transitions of both types. They are standard models such as Ising model, lattice gas model, XY
model \cite{lit1} but also some innovatory ones \cite{lit2}. The initial point for most of them is Hamiltonian of the system for which partition
function (or another statistical quantity) can be calculated. Then the partition function is examined and discontinuities or singularities,
which may be responsible for the presence of the phase transition, are searched \cite{lit3}.

In our model the situation is opposite to some extent. We start constructing the model on the intuitive premise and only then the formal analysis
may reveal that the partition function of the system has singularity at the thermodynamic limit .

What features of crystal-liquid or liquid-gas transitions, apart from the latent heat, are most characteristic? It can be said that it is a well
defined temperature (for fixed pressure) of the transition. Below this temperature we have only phase I (e.g. crystalline solid) while above there
is only phase II (e.g. liquid). Both phases (generally) differ considerably from each other (symmetry, viscosity, compressibility, etc.), what means
that a very little change of temperature of the system causes a dramatic change in the arrangement of molecules. Main purpose of our model is to explain
intuitively how it is possible.

A process of random walking at discrete time was adopted in our model in order to describe dynamics. Basic properties of melting are shown to be caused
by the fact that there exists a border between the phases. Only near that border molecules may change their attachment to the phase. Far off the border
they are blocked by their neighbours. Such a situation results in an exponential distribution of the energy spectrum of the system, the fact being a
condition for the relation between energy and entropy to be linear so that phase transition may take place.

\section{Description of the Model}

\subsection{The Basic Element of the Model}

The quantum system with a free hamiltonian $H_0$ is a basic element of our model. The spectrum consists of $n+m+1$ (discrete, non-degenerated) levels,
which are shown in Fig.~1. Such a system is called a molecule.

The lower $n+1$ levels, numbered from 0 to $n$, are called the levels of the first phase while the upper $m$ levels (from $n+1$ to $n+m+1$) are the
levels of the second phase.

\begin{figure}[h]
\begin{center}
\includegraphics[width=4.5cm]{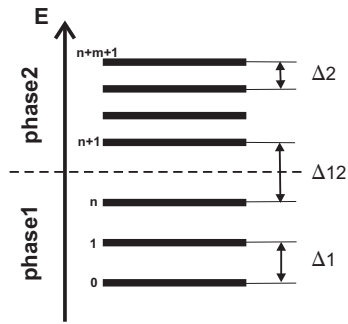}
\end{center}
\caption{Spectrum of a single molecule}
\end{figure}

Energy of the $k$-th level of the molecule is given by:
\begin{displaymath}
E_k=\left\{ \begin{array}{ll}
\Delta_1\,k & \textrm{for $k \leq n$}\\
\Delta^*_{12}+\Delta_2(k-n-1) & \textrm{for $k>n$},
\end{array} \right.
\end{displaymath}
where $\Delta_{12}^*=\Delta_{12}+\Delta_1\,n$ is energy of lowest level of phase II and $\Delta_{12}$ is energy 
gap between the phases, the existence of with is not necessary but added to make our consideration more general.
What is more, our model is operational even if we assume $\Delta_1=\Delta_2=\Delta_{12}$.

Neither the evident form of $H_0$ nor its ,,origin'' is needed. However, it is better for intuitive understanding to imagine that the states
of $H_0$ are the states of the motion of a molecule rather  than  the states of its internal electron excitations. 

In the following part of our work we examine behaviour of such molecules which, due to the interactions with external environment (thermostat),
change their state and jump at random to a neighbouring level (upper or lower). The chance of jumping has been limited to transitions between
the neighbouring levels only. Such a limitation is merely technical simplification and does not affect the appearance of the phase transition.

\subsection{Scheme of the System and Rules of Transition}

In the present paper we examine a system consisted of $N$ molecules (as introduced in the previous subsection), which are numbered from 1 to
$N$ and set into a one-dimensional lattice (Fig.2).

\begin{figure}[h]
\begin{center}
\includegraphics[width=6cm]{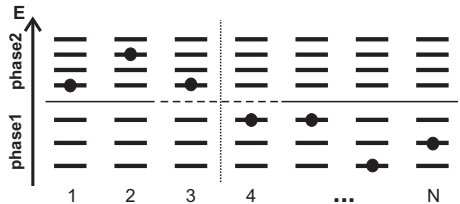}
\end{center}
\caption{Scheme of the lattice}
\end{figure}

Additionally, it was assumed that individual molecules can interact with each other. The interaction does not change the structure
of levels in Fig.~1 but causes that some of them become unavailable for the molecule, what means that there appear some restriction
rules (which are essential for the phase transition to come into being):

\begin{enumerate}
\item the $i^{th}$ molecule may jump up from the top level of the first phase (i.e. $n^{th}$ level) onto the lowest level of the second
phase (i.e. $(n + 1)^{th}$ level) only if the left (i.e. $(i-1)^{th}$) neighbour of the molecule  is currently in the second phase,

\item the $i^{th}$ molecule may jump down off the lowest level of the second phase (i.e. $(n + 1)^{nt}$ level) to the top level of the
first phase ($n^{th}$ level) only if the right ($(i + 1)^{th}$) neighbour of the molecule is currently in the first phase,

\item exceptions: first molecule ($i=1$) is always able to jump from $n^{th}$ level to $(n+1)^{th}$ level,
 last molecule ($i=N$) is always able to jump from $(n+1)^{th}$ level to $n^{th}$ level,

\item only that distribution of states was always chosen as the initial state of the simulation, for which all the molecules
in phase II lie on the left to all of the molecules in phase I.
\end{enumerate}

The foregoing rules set up a border between the phases (analogous to a free surface) in the system. Transitions of
molecules from the first phase to the second phase are possible only at the border of the phases. A molecule, which gathers
enough energy to jump into a different phase, is not able to do it until its appropriate neighbour has done it first so the
molecule is blocked by its neighbour and has to wait for its turn. It is shown below that such a property of the model guarantees
that the first-order phase transition can exist.

\subsection{Limitations on Our Model}

\begin{itemize}
\item our model is one-dimensional while systems in nature are usually 3-dimensional,

\item molecules in phase II are distinguishable in our lattice while in gases and liquids they are undistinguishable,

\item molecules in phase I have energy levels which are mutually independent in our model while in crystals their energy
levels are common (collective fonon excitations),

\item distribution of levels in both phases of our model are uniform; there are regular gaps between the levels, 

\item in our model the space, which is occupied by each phase is compact (i.e. it is in one piece) while in the real systems
the phases may by mixed with each other as in fog, for example, when one phase is a suspension in another phase. Such an
assumption is of a special kind because it is not connected with a character of local interaction between molecules but
introduces some global limitations to possible configurations of the system. Although non-natural, the assumption is necessary
for mathematical simplifications. Our model is only a simple demonstration of process of growing (decline) a single drop of
liquid during condensation (evaporation).
\end{itemize}

\section{Theoretical Analysis of the Properties of the Model}

\subsection{Statistical Properties of a Single Molecule}

Let us consider a single (free\footnote{i.e. without limitations from section 2.2}) molecule defined in Fig.~1.
A canonical partition function for such a system is given by:
\begin{equation}
Z_{n,m}=A_n+B_{n,m}
\end{equation}
where $A_n$ is contribution of the states, which belong to phase I:
\begin{equation}
A_n=\sum_{k=0}^{n}\exp(-\beta\Delta_1k)=\frac{1-\exp(-\beta\Delta_1(n+1))}{1-\exp(-\beta\Delta_1)},
\end{equation}
while $B_{n,m}$ is a contribution of states, which belong to phase II:
\begin{equation}
B_{n,m}=\sum_{k=n+1}^{n+m}\exp(-\beta(\Delta^*_{12}+(k-n-1)\Delta_2))=
\frac{\exp(-\beta\Delta_{12}^*)(1-\exp(-\beta m \Delta_2))}{1-\exp(-\beta\Delta_2)}.
\end{equation}

\noindent In low temperature (i.e. for $A_n>B_{n,m}$) a free molecule can be found in the first phase more often than in the second.
If $m>n+1$ (i.e. number of levels in phase II is greater than the number of levels in phase I) the opposite situation is
possible. In high temperatures (when $A_n<B_{n,m}$) the probability that a free molecule is in the second phase is greater than the
probability of finding it in first phase. There is a temperature at which the levels of both phases can be occupied by molecules
with the same probability:
\begin{equation}
\label{eq4}
A_n(\beta^*)=B_{n,m}(\beta^*).
\end{equation}
It will be shown that such a temperature is the one of the phase transition in our model (Fig.~2).
This temperature will be marked by $T^*$ and its reverse by $\beta^*$.

\subsection{Statistical Properties of the Lattice}

Let us introduce Hamiltonian of the lattice (Fig.~2) in an evident form:
\begin{equation}
H=\oplus_{i=1}^N H_0 + H_{int}
\end{equation}
where $\oplus$ is direct sum, $H_0$ is a free Hamiltonian of single molecule, and $H_{int}$ is an interaction Hamiltonian
responsible for restriction rules as given in section 2.2. The way hamiltonian $H_{int}$ acts can be imagined as that: it
increases energies of ,,forbidden'' eigenstates of hamiltonian $\oplus_{i=1}^N H_0$ by a certain very high value $\Omega$
what results in an insignificant probability of forbidden eigenstates to be occupied. 

\noindent
Partition function of the lattice is given by:
\begin{equation}
\label{eqzl}
z_l(\beta)=\sum_{i=0}^NB_{n,m}^iA_n^{N-i}
\end{equation}
where $A_n$ and $B_{n,m}$ are given by Eq.~(2) and (3).
The formula~(\ref{eqzl}) can be written in a following form:
\begin{multline}
z_l(\beta)=\theta(\beta-\beta^*)\,A_n^N\sum_{i=0}^Na^i+\theta(\beta^*-\beta)\,B_{n,m}^N\sum_{i=0}^N(a^{-1})^i=\\
=\theta(\beta-\beta^*)A_n^N\frac{1-a^{N+1}}{1-a}+
\theta(\beta^*-\beta)B_{n,m}^N\frac{1-(a^{-1})^{N+1}}{1-a^{-1}}
\end{multline}
where $a=B_{n,m}/A_n$, and $\theta$ is a Heaviside function.

The function $z_l$ can be given in a form of two components, therefore a limit can be calculated for each
of them as $N\rightarrow\infty$. It can be seen that $z_l$ function gets divergent as $N$ approaches infinity
(due to $A^N$ and $B^N$). However, with $\beta$ tending to $\beta^*$ there appears a further divergence
(divergence of the sum of geometric series).

\noindent
Averaged energy of the lattice per single molecule is given by:
\begin{multline}
\label{eqtott}
E(\beta)=-\frac{1}{N}\,\frac{\partial\textrm{ln}z_l}{\partial\beta}=
\theta(\beta-\beta^*)\biggl(-\frac{\partial\textrm{ln}\,A_n}{\partial\beta}-
\frac{1}{N}\frac{\partial}{\partial\beta}\textrm{ln} \frac{1-a^{N+1}}{1-a} \biggr)+\\
+\theta(\beta^*-\beta)\biggl(-\frac{\partial\textrm{ln}\,B_{n,m}}{\partial\beta}-
\frac{1}{N}\frac{\partial}{\partial\beta}\textrm{ln} \frac{1-(a^{-1})^{N+1}}{1-a^{-1}} \biggr).
\end{multline}
For $N\rightarrow\infty$ we obtain:
\begin{equation}
\label{eq15}
E(\beta)=\theta(\beta-\beta^*)\, \langle{E_I}\rangle+\theta(\beta^*-\beta)\,\langle{E_{II}}\rangle,
\end{equation}
where $\langle{E_I}\rangle$ and $\langle{E_{II}}\rangle$ are average energies of molecule in the first and the
second phase respectively.

\noindent
For phase I (when levels of phase II are unavailable) averaged energy is given by:
\begin{equation}
\label{eq12}
\langle{E_I}\rangle_{(\beta)}= -\frac{\partial\textrm{ln}A_n}{\partial\beta}=
\Delta_1\biggl(\frac{1}{\exp(\beta\Delta_1)-1}-\frac{(n+1)}{\exp(\beta\Delta_1(n+1))-1}\biggr)
\end{equation}
while in phase II (when phase I is unavailable) it is given by:
\begin{equation}
\label{eq13}
\langle{E}_{II}\rangle_{(\beta)}= -\frac{\partial\textrm{ln}B_{n,m}}{\partial\beta}=
\Delta_{12}^{*}+\Delta_2\biggl(\frac{1}{\exp(\beta\Delta_2)-1}-\frac{m}{\exp(\beta\Delta_2m)-1}\biggr).
\end{equation}

\noindent
It can be seen that for $N\rightarrow\infty$ an energy leap takes place at temperature $T=T^*$ (see eq.~\ref{eq15}),
what let us define latent heat per single molecule in the following form:
\begin{equation}
\label{eq14}
c_T=\lim_{\beta\rightarrow\beta^{*-}}E(\beta)-\lim_{\beta\rightarrow\beta^{*+}}E(\beta)
=\langle{E}_{II}\rangle_{(\beta^*)}-\langle{E_I}\rangle_{(\beta^*)}.
\end{equation}

\noindent
In the state of equilibrium a probability of finding $k$ molecules in phase II is given by:
\begin{equation}
\label{eq111}
P_k=\frac{B_{n,m}^k\,A_n^{N-k}}{z_l}.
\end{equation}
It is worth noticing that:
\begin{equation}
\label{eqrr}
P_{k+1}=a\,P_k,
\end{equation}
what allows (when taken into account that probabilities are normalised to unity) for eq.~(\ref{eq111}) to be
written in a following form:
\begin{equation}
\label{eq16}
P_k=\frac{1-a}{1-a^{N+1}}a^k,
\end{equation}
where $a=B_{n,m}/A_n$.\\
With the use of the formula~(\ref{eq16}) an average number of molecules occupying phase II can be calculated
in the form:
\begin{equation}
N_{II}=\sum_{k=0}^Nk\,P_k=\frac{a\,(1+N\,a^{N+1}-a^N(1+N))}{(1-a)(1-a^{N+1})}
\end{equation}
and average number of molecules occupying phase I is defined as $N_I=N-N_{II}$.

\noindent
An approximate formula for energy of the lattice per single molecule as a function of temperature is given by:
\begin{equation}
\label{eq18}
\tilde{E}(\beta)=(N_I\,\langle{E_I}\rangle+N_{II}\,\langle{E_{II}}\rangle)\,N^{-1},
\end{equation}
where averaged energies are given by eq.~(\ref{eq12}) and eq.~(\ref{eq13}).
Formula (\ref{eq18}) is an estimation as it was not considered that near the border between the phases molecules
had not enough time to reach the equilibrium state in their current phase (i.e. molecules did not ''forget'' yet
that some time ago they had occupied the other phase).

\subsection{Connection with Random Walking}

It is convenient to use Markov chains to make it possible to understand why phase transition occurs in our model.
The states of the process are numbered 0 to $N$ and they define the number of molecules occupying phase II
(or in other words a position of the phase border in our lattice).

\noindent That process is a random walking (Fig.~3), with probabilities of jumping right and left given by:
\begin{eqnarray}
\label{eq5}
\alpha=\frac{B_{n,m}}{A_n+B_{n,m}}, \\
\label{eq6}
\beta=\frac{A_n}{A_n+B_{n,m}},
\end{eqnarray}
where $\alpha$ - probability of transition of the next molecule to phase II, $\beta$ - probability of transition of
the next molecule to phase I. Probabilities (\ref{eq5}), (\ref{eq6}) are chosen so that $\alpha+\beta=1$ and a
stationary distribution of Markov chain would be in accordance with the equilibrium distribution (\ref{eq111})-(\ref{eq16}).

\begin{figure}[h]
\begin{center}
\includegraphics[width=6cm]{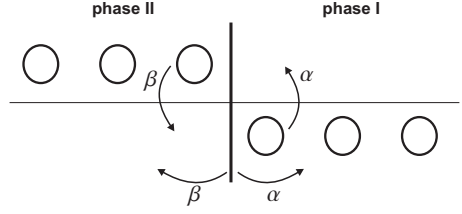}
\end{center}
\caption{Markov chain of transition of molecules between the phases}
\end{figure}

\noindent
Referring to well known properties of random walking \cite{lit4} we get the following facts:

\begin{itemize}
\item for $T<T^*$ (i.e. for $\alpha<\beta$) stationary distribution of the process is exponential (left plot in Fig.~4),
 what means that in the state of equilibrium most of the molecules occupy levels of phase I. Only at the left edge of
 the lattice there are some fluctuations leading to the fact that some molecules occupy levels of phase II.
 With $T$ approaching $T^*$ fluctuations grow stronger and stationary distribution tends to become uniform,

\item for $T>T^*$ (i.e. for $\alpha>\beta$) the situation is opposite (right plot in Fig.~4). Majority of the molecules
 are in phase II except for fluctuations at the right edge of the lattice which cause that some molecules occupy levels
 of phase I,

\item when $T=T^*$, it is a case of symmetrical random walking ($\alpha=\beta$), which corresponds to a uniform distribution
 of probability $p_n=1/N$. Large fluctuations of energy are produced during the phase transition, what is apparent if we
 remind an example of the system that consists of the mixture of water and ice, at temperature $0^oC$, and set in a thermostat
 containing water and ice. A canonical distribution is used there, and fixed temperature in the zone of phase transition
 (with latent heat) does not establish the energy of the system.

\end{itemize}

\noindent
For a very large lattice (when $N\rightarrow\infty$) the effect of edge fluctuations on all the properties of the system is
negligible, even if $T\approx T^*$ (i.e. $\alpha\approx\beta$). With  the  increase of temperature of the system it can
be noticed that nearly all the molecules change their phase simultaneously when temperature is exceeding $T^*$.
Such a behaviour could be spotted in practice, in numerical simulations described in the next sections.

\begin{figure}[h]
\begin{center}
\includegraphics[width=5cm]{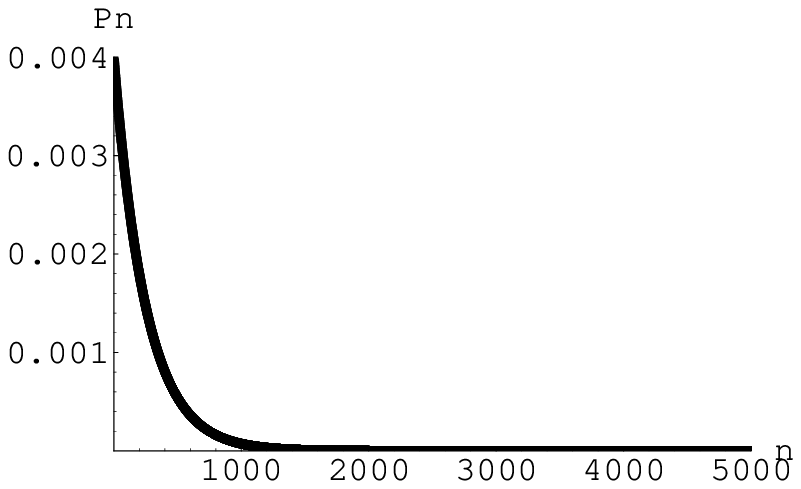}
\includegraphics[width=5cm]{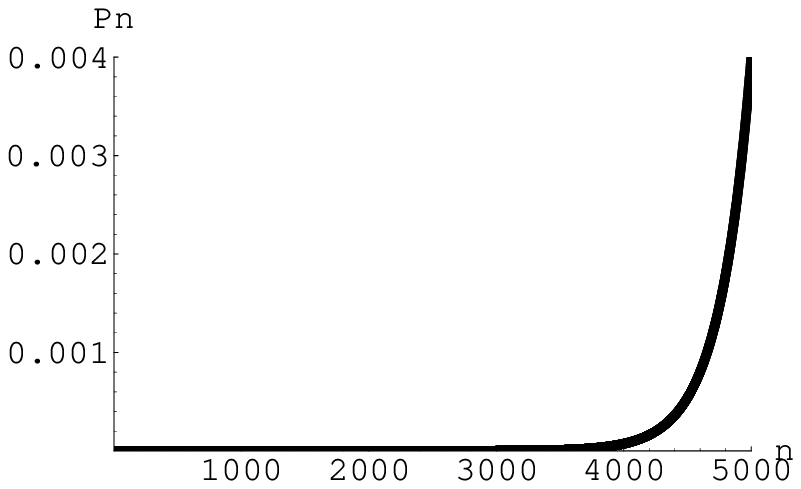}
\end{center}
\caption{Stationary distribution for asymmetrical random walking $N=5000$ (left figure: $\alpha=0.501$, $\beta=0.499$,
right figure: $\alpha=0.499$, $\beta=0.501$)}
\end{figure}

In our model, a transition of a molecule from phase II to phase I is possible only if it joins itself to another molecule
that is already occupying phase I (with except for the first molecule, which may always jump to phase I).
It is in correspondence to real solidification, where molecules most often join an already existing crystal or begin
to crystallise on microscopic grains of dust. In our model the right end of the lattice plays a role of a grain, because
molecule $i=N$ can always jump to phase I. Had that rule been eliminated from our model, all molecules being converted
to phase II in high temperature, then after another fall of temperature molecules would not be able to return to phase I
what would produce a supercooled phase II.

\subsection{Examples}

\begin{figure}[h]
\begin{center}
\includegraphics[width=6cm]{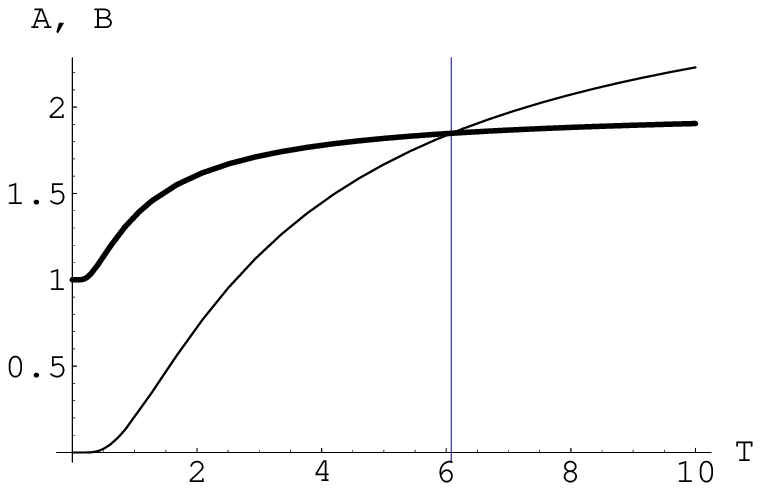}
\includegraphics[width=6cm]{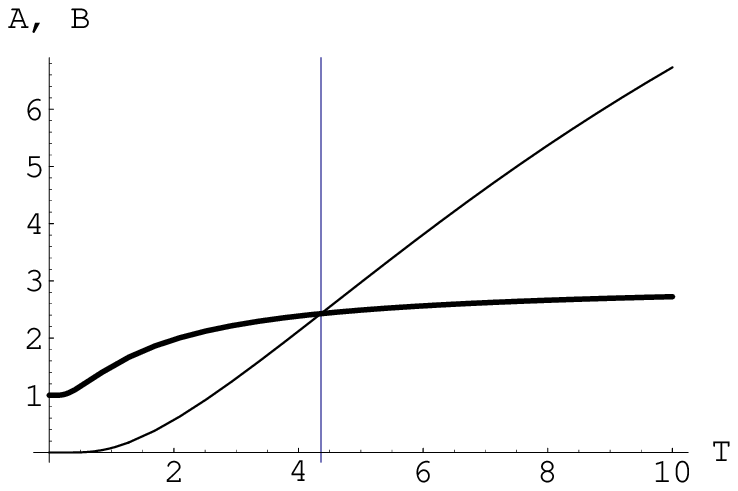}
\end{center}
\caption{Calculation of transition temperature $T^{*}$. Left plot: example 1 (section 3.4.1), right plot: example 2
(section 3.4.2)}
\end{figure}

\begin{figure}[h]
\begin{center}
\includegraphics[width=6cm]{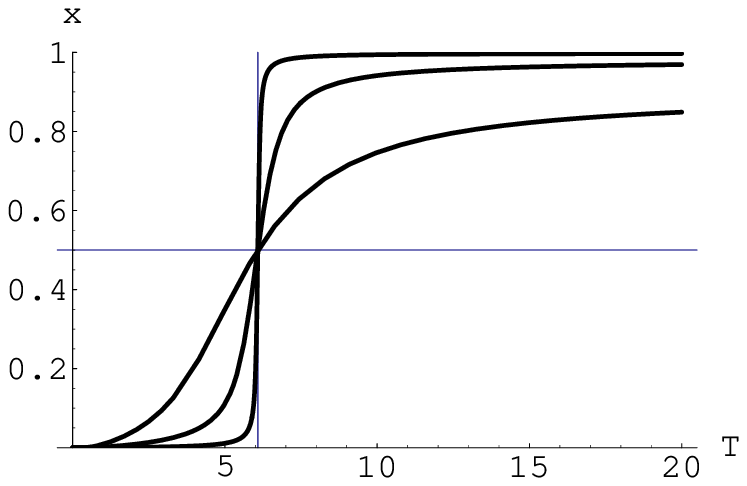}
\includegraphics[width=6cm]{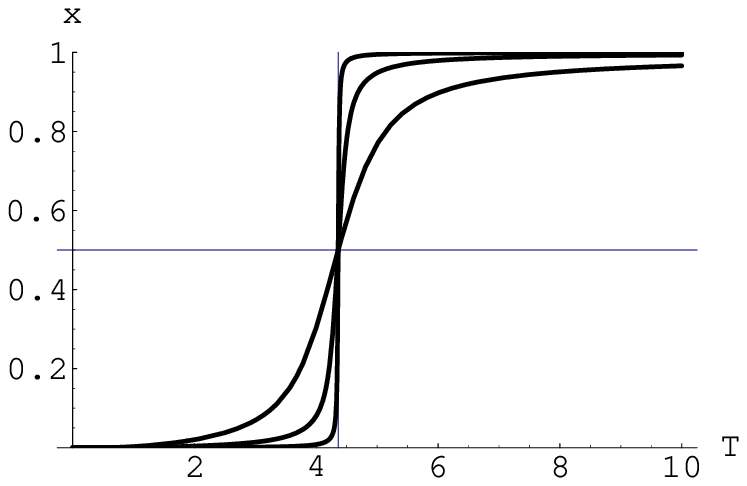}
\end{center}
\caption{Amount of molecules occupying phase II ($x=N_{II}/N$) as function of temperature (curves for $N=20, 100, 1000$).
Left plot: example 1, right plot: \mbox{example~2}}
\end{figure}

\begin{figure}[h]
\begin{center}
\includegraphics[width=6cm]{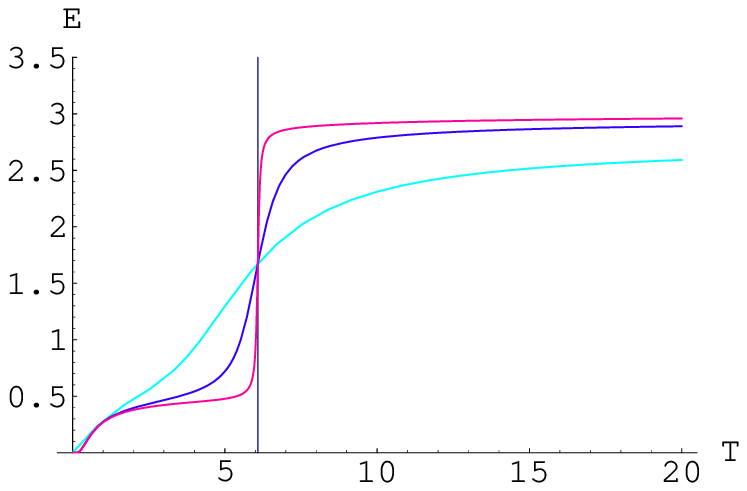}
\includegraphics[width=6cm]{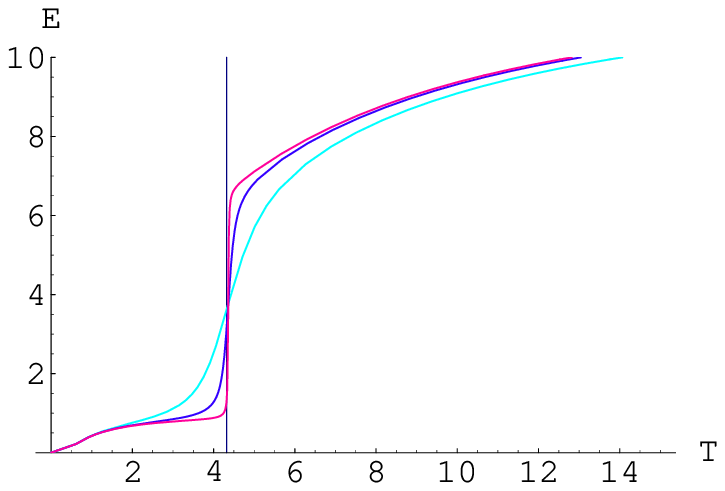}
\end{center}
\caption{Average energy of the lattice per single molecule as function of temperature (curves for $N=20, 100, 1000$).
Left plot: example 1, right plot: \mbox{example~2}}
\end{figure}

\subsubsection{Example 1}
Let $n=1$, $m=3$, $\Delta_1=\Delta_2=\Delta_{12}=1$ then:
\begin{itemize}
\item temperature of transition, calculated according to Eq.(\ref{eq4}) with the use of a numerical method
 (Fig.~5 - left plot) is $T^{*}=6.08152$,
\item average energy of the molecule in phase I for $T=T^*$ calculated according to Eq.(\ref{eq12}) is
$\langle E_I\rangle_{(\beta^*)}=0.458984$,
\item average energy of the molecule in phase II for $T=T^*$ calculated according to Eq.(\ref{eq13}) is
$\langle E_{II}\rangle_{(\beta^*)}=2.89087$,
\item latent heat per single molecule calculated basing on Eq.(\ref{eq14}) is $c_T=2.431886$.
\end{itemize}

\subsubsection{Example 2}
Let $n=2$, $m=20$, $\Delta_1=\Delta_2=\Delta_{12}=1$ then:
\begin{itemize}
\item temperature of transition, calculated according to Eq.(\ref{eq4}) with the use of numerical method (Fig.~5 right plot)
 is $T^{*}=4.3602$,
\item average energy of the molecule in phase I for $T=T^*$ (calculated according to Eq.(\ref{eq12})) is
 $\langle E_I\rangle_{(\beta^*)}=0.848427$,
\item average energy of the molecule in phase II for $T=T^*$ (calculated according to Eq.(\ref{eq13})) is
 $\langle E_{II}\rangle(T^*)=6.67352$,
\item latent heat per single molecule calculated basing on Eq.(\ref{eq14}) is $c_T=5.825093$.
\end{itemize}

\noindent
The plots in Fig.~6 depict the amount of molecules in phase II ($x=N_{II}/N$) against temperature.
The left plot represents the example 1 while the right - example 2.
The plots in Fig.~7 depict average energy of the system per single molecule (Eq.~\ref{eqtott}) as function of temperature
(right plot for example 1, left plot for example 2).
We can see that with increasing $N$ the curves become more and more steeper near the temperature of the phase transition.

\section{Numerical Simulation}

\subsection{Monte-Carlo Simulation}

In order to verify our consideration in practice we created a computer program (simulation in Java language).
We examined lattices, which were described in examples 1 and 2 (last section) with the use of Monte Carlo algorithm\cite{lit5}.
In our algorithm, each molecule changes its position at random to higher or lower at each step and probability of such a leap
depends on temperature as follows:
\begin{eqnarray}
A&=&\frac{\exp(-\beta\Delta)}{1+\exp(-\beta\Delta)},\\
B&=&\frac{1}{1+\exp(-\beta\Delta)},
\end{eqnarray}
where $A$ - probability of jumping up to a neighbouring higher level (if transition is permitted),
$B$ - probability of jumping down to a neighbouring lower level (if transition is permitted),
$\Delta$ - energy gap between the levels.

\begin{figure}[h]
\begin{center}
\includegraphics[width=6cm]{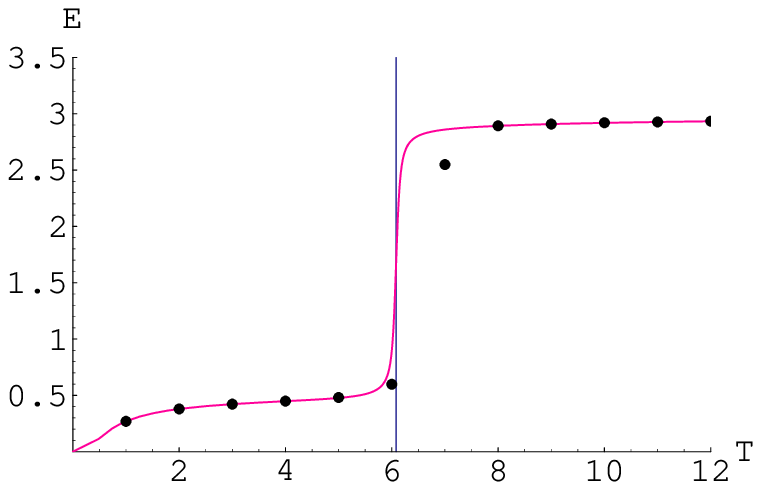}
\includegraphics[width=6cm]{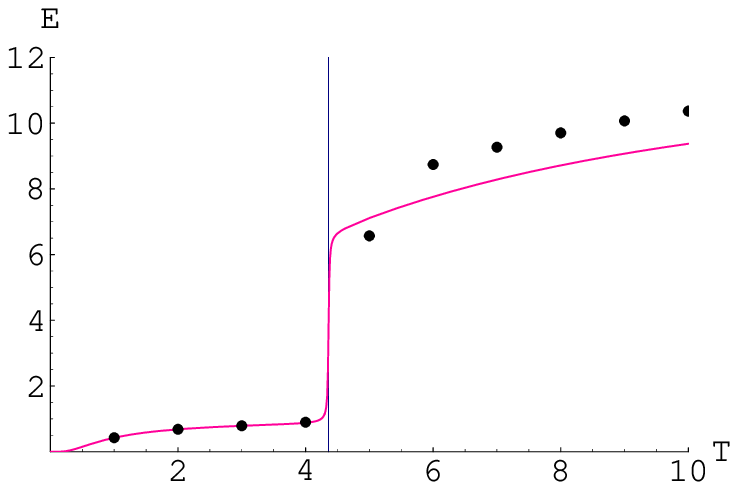}
\end{center}
\caption{Average energy of the lattice per molecule as function of temperature for $N=1000$ (left plot: first example,
right plot: second example)}
\end{figure}

The results of our simulation are shown in figure~8. Left plot is drawn for the first example (section 3.4.1), right for
the second one (section 3.4.2).  The solid line is added for comparison - it is a theoretical curve (\ref{eqtott}),
formerly shown in Fig.~7. Fat dots are end products of numerical simulation. Each of them represents (averaged over time)
energy of the lattice (per a single molecule), for 500\,000 time steps. 

\noindent Simulation in the form of Java applet is available at www...(soon)

\subsection{Numerical Computation of Distributions of Levels and of Other Thermodynamic Properties}

For small lattices (i.e. for small $N$, $m$ and $n$) it is possible to calculate energy for each state of the
lattice directly (numerically). It was done with the use of a computer program for the system described in the
first example (section 3.4.1) and for $N=20$ molecules.
We have obtained:\\
- 81 energy levels of the system (first level: $E=0$, last: $E=80$),\\
- 10\,458\,256\,051 states of the system of all,\\
- degeneracy $d_n$ of energy levels (i.e. number of states of the system with the same energy) - see appendix A.
  In Fig.~9 logarithm $d_n$ is plotted against energy $E_n$\\

\begin{figure}[h]
\begin{center}
\includegraphics[width=6cm]{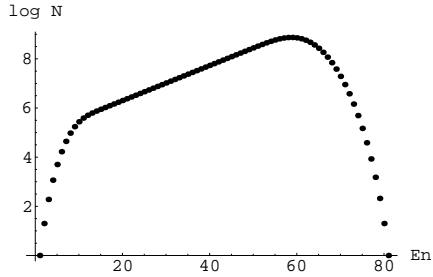}
\end{center}
\caption{Energy levels degeneracy $d_n$ (in logarithmic scale) as function of energy}
\end{figure}

In the Fig.~9 a linear fragment of the curve can be noticed, which corresponds to exponential increase of level degeneracy.
If degeneracy $d_n$ was identified with density of energy levels $\rho(E)$ and $S=k\,\ln\rho(E)$ then $S(E)$ would be
linear and responsible for phase transition occurrence.

\noindent
It can be proved intuitively in a following way that density of levels increases exponentially:
\begin{itemize}
\item Let $D_i$ be amount of states of the system, which are available when border between the phases is located near $i$-th
 molecule (i.e. $N_{II}=i$). Then $D_{i+1}=D_i\,m/(n+1)$, what means that $D_i$ increases exponentially with respect to the number
of molecules occupying phase II ($N_{II}$),
\item refering to Eq.~(\ref{eq18}) we can see that energy of the system during the phase transition ($T=T^*$) is linear
function of $N_{II}$,
\item both above-mentioned facts joined togehter let us conclude that during the phase transition density of energy levels should
increase exponentially with respect to energy of the system.
\end{itemize}

\begin{figure}[h]
\begin{center}
\includegraphics[width=10cm]{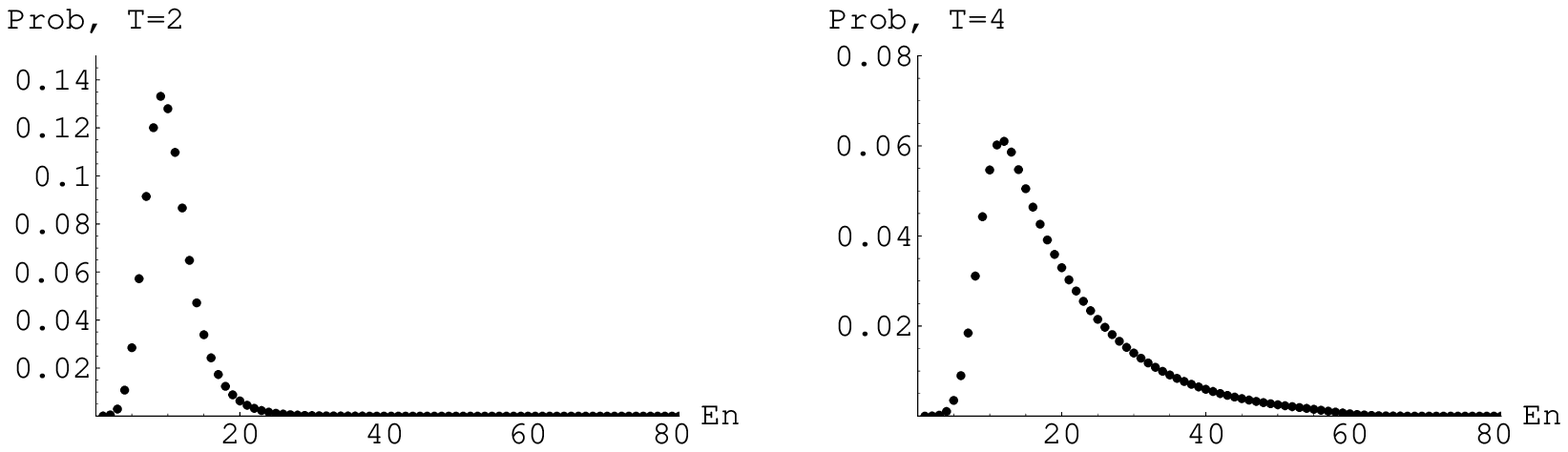}
\includegraphics[width=10cm]{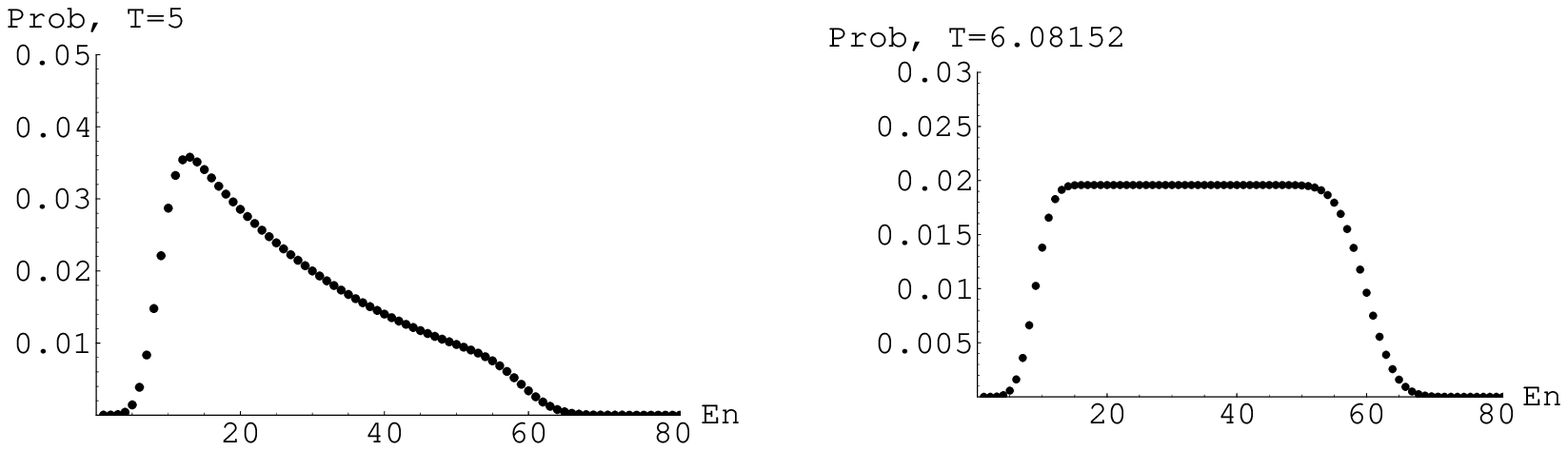}
\includegraphics[width=10cm]{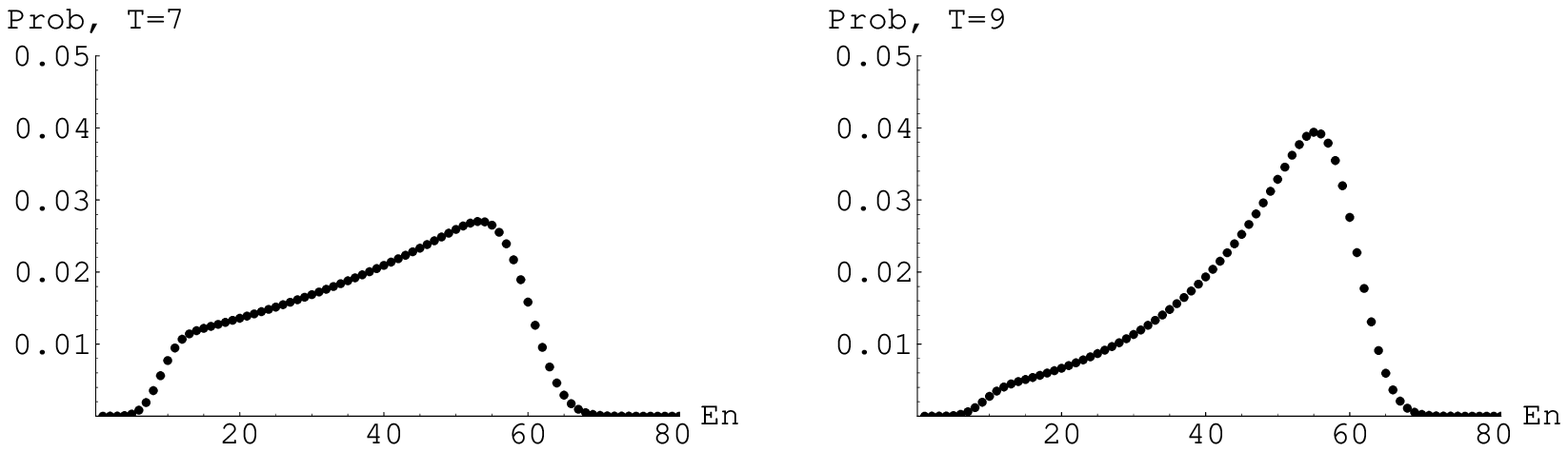}
\caption{Distribution of probability of level occupation for temperatures $T$=2, 4, 5, $T^*$, 7, 9}
\end{center}
\end{figure}

\noindent
When numbers $d_n$ are known, they can be used at calculation of probability distribution of levels to be occupied at
fixed temperature with the use of the formula:
\begin{equation}
\label{eq22}
p_n(\beta)=\frac{d_n}{Z_{tot}}\exp(-\beta E_n).
\end{equation}
In Fig.~10 the distribution (\ref{eq22}) is plotted for several temperatures, the temperature of the transition being
also included. It is worth noticing that the distribution for $T=T^*$ gets flat (as it was expected) since there are
symmetrical random walking and large energy fluctuations then.

To compare numerical results with theoretical prediction (according to section~3) an average energy of the system
(per a single molecule) was calculated as follows:
\begin{equation}
\langle{E}\rangle(\beta)=\sum_nE_n\,p_n,
\end{equation}
and is shown versus temperature (fat curve in Fig.~11), where $p_n$ are given by Eq.(\ref{eq22}) and energies are equal
to $E_n=n\Delta$. The hardly visible thin curve in Fig.~11 (which almost merges with the thick one) represents the
theoretically computed energy (as in Eq.~(\ref{eqtott})) and was alredy presented earlier in the left plot of Fig.~7.
We can see that numerical results and theoretical expectation of the last section are in accordance.

\begin{figure}[h]
\begin{center}
\includegraphics[width=5cm]{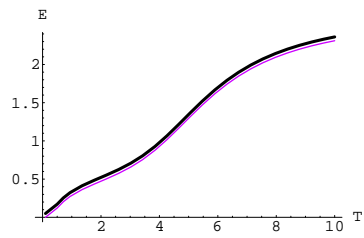}
\caption{Average energy of the system per single molecule versus temperature (comparison)}
\end{center}
\end{figure}

\newpage
\section*{Acknowledgements}
I wish to thank P.~Kosinski, B.~Broda for discussions about results; E.~Zujewska for language help.

\section*{Appendix A. Degeneracies of energy levels of the system}

$d_n$=\{1, 20, 191, 1160, 5037, 16683, 43969, 95191, 173988, 275785, 390079, 507601, 626181, 750774, 889367, 1049453,
1237218, 1458361, 1719007, 2026235, 2388372, 2815232, 3318382, 3911457, 4610529, 5434542, 6405826, 7550702, 8900195,
10490875, 12365848, 14575924, 17180994, 20251653, 23871113, 28137458, 33166302, 39093922, 46080951, 54316731, 64024444,
75467160, 88954965, 104853349, 123592989, 145680735, 171710383, 202368411, 238422447, 280667788, 329792568, 386115323,
449167565, 517155934, 586442932, 651295360, 704192361, 736884109, 742127143, 715675486, 657851392, 574032176, 473729446,
368478879, 269241736, 184195595, 117577731, 69769278, 38326383, 19398465, 8996217, 3797379, 1447096, 492785, 147992,
38513, 8475, 1521, 210, 20, 1\}\\

\noindent Numbers listing in the bracket are degeneracies of energy levels of the lattice
(from $E=0$ to $E=80$) - see section~4.2.

\end{document}